\documentclass[pra,twocolumn,superscriptaddress,amsmath,amssymb]{revtex4}
\usepackage{graphicx}
\usepackage{bm}
\usepackage{amsmath,amssymb}
\usepackage{color}
\include{Qcircuit}

\begin{document}
\title{Classical simulatability of the one clean qubit model}
\author{Tomoyuki Morimae}
\email{morimae@gunma-u.ac.jp}
\affiliation{ASRLD Unit, Gunma University, 1-5-1 Tenjin-cho Kiryu-shi
Gunma-ken, 376-0052, Japan}
\author{Takeshi Koshiba}
\affiliation{Graduate School of Science and Engineering, Saitama University,
255 Shimo-Okubo, Sakura, Saitama 338-8570, Japan}

\date{\today}
            
\begin{abstract}
Deterministic quantum computation with one quantum bit (DQC1),
or the one clean qubit model,
[E. Knill and R. Laflamme, Phys. Rev. Lett. {\bf81}, 5672 (1998)]
is a model of quantum computing
where the input is the tensor product of a single pure qubit and
many completely-mixed states,
and only the single qubit is measured at the end of the computation.
In spite of its naive appearance,
the DQC1 model can efficiently solve
some problems for which no classical efficient algorithms are known,
and therefore it has been conjectured that
the DQC1 model is more powerful than
classical computing (under the assumption
of $\mbox{BPP}\subsetneq\mbox{BQP}$).
However, there has been no proof for the conjecture.
Here we show that 
the output probability distribution of the DQC1 model
cannot be classically efficiently approximated
(exactly within a polynomial bit length
or
in the fully polynomial randomized approximation scheme (FPRAS)
with at most a constant error)
unless $\mbox{BQP}\subseteq\mbox{BPP}$.
\end{abstract}

\maketitle  

\section{Introduction}
\begin{figure}[htbp]
\begin{center}
\includegraphics[width=0.45\textwidth]{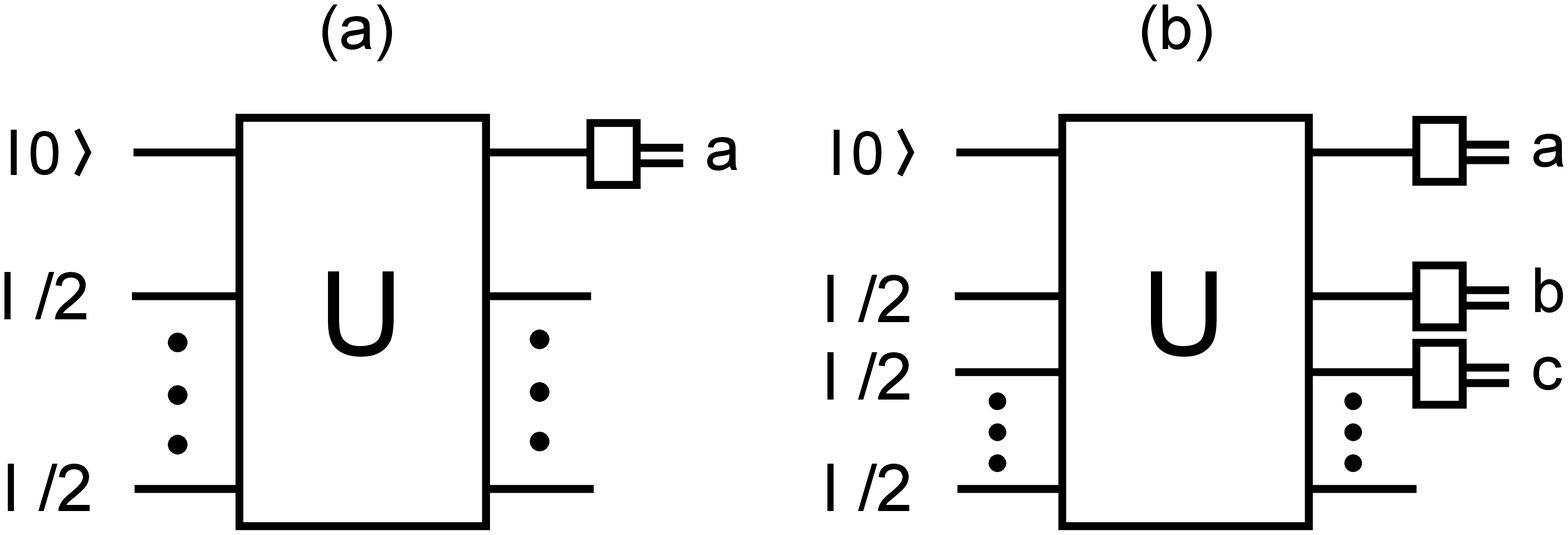}
\end{center}
\caption{(a) The DQC1 model. (b) The DQC1$_m$ model for $m=3$.
} 
\label{DQC1}
\end{figure}

The deterministic quantum computation with one quantum bit (DQC1),
or the one clean qubit model,
proposed by Knill and Laflamme~\cite{KL} is a restricted model
of quantum computing. It was originally motivated by nuclear magnetic resonance (NMR) 
quantum information processing, but now it has been extensively studied
in variety of contexts
to understand the border between quantum and classical 
computing~\cite{Poulin,Poulin2,SS,Passante,JW,3mani,Datta,Datta2,Datta3,
MFF,Elham}.
As shown in Fig.~\ref{DQC1} (a), a DQC1 circuit consists of 
\begin{itemize}
\item[1.]
the input state $|0\rangle\langle0|\otimes\frac{I^{\otimes n}}{2^n}$,
where $I\equiv|0\rangle\langle0|+|1\rangle\langle1|$
is the two-dimensional identity operator,
\item[2.]
a $poly(n)$-size quantum gate $U$,
\item[3.]
the computational basis measurement of the first qubit,
which gives the output bit $a\in\{0,1\}$. 
\end{itemize}

The DQC1 model seems to be very weak, and in fact,
it does not support universal quantum computation
under reasonable assumptions~\cite{Ambainis}. 
However, counter-intuitively, the DQC1 model 
can efficiently solve some problems for which no efficient classical algorithm is 
known, such as the spectral density estimation~\cite{KL}, 
testing integrability~\cite{Poulin}, 
calculation of fidelity decay~\cite{Poulin2},
approximation of the Jones and HOMFLY polynomials~\cite{SS,Passante,JW}
and an invariant of 3-manifolds~\cite{3mani}.
These results are surprising since if we replace the single pure input qubit
with the completely mixed qubit, then the output suddenly becomes trivially classically
simulatable. In other words, the single pure qubit hides some capacity
of unlocking the potential power of highly-mixed states
for the quantum speedup.
In fact, such an unexpected power of the DQC1 model was shown to
come from some non-classical correlations
between the mixed register and the pure qubit~\cite{Datta,Datta2,Datta3}.
In short, the DQC1 model is believed 
to be a model of computation which is intermediate classical 
and universal quantum computation.

In Ref.~\cite{MFF},
a generalized version of
the DQC1 model, which is called the DQC1$_m$ model,
was introduced.
The DQC1$_m$ model is the same as the DQC1 model except
that not a single but $m$ output qubits are measured
in the computational basis
at the end of the computation (Fig.~\ref{DQC1} (b)).
In particular, the DQC1$_1$ model is equivalent to the DQC1 model.
It was shown in Ref.~\cite{MFF} that the output probability
distribution of the DQC1$_m$ model
for $m\ge3$
cannot be classically
efficiently sampled (within a multiplicative error) unless the polynomial hierarchy 
collapses at the third level~\cite{MFF}. 
The polynomial hierarchy~\cite{Toda} is a natural way of classifying the complexity 
of problems (languages) beyond the usual NP (nondeterministic polynomial time,
which includes ``traveling salesman" and ``satisfiability" problems).
Since it is believed in computer science that 
the polynomial hierarchy does not collapse, 
it is unlikely that the DQC1$_m$ model for $m\ge3$ can be classically efficiently
sampled. This result is based on the ``postselection technique"
used for other restricted models of quantum computing,
such as the depth-four circuits model by Terhal and DiVincenzo~\cite{TD},
the instantaneous quantum polytime (IQP) model by Bremner, Jozsa, and Shepherd~\cite{BJS}, and
the Boson sampling model by Aaronson and Arkhipov~\cite{AA}.

Can we show an impossibility
of classical efficient simulation of the DQC1$_m$ model
for $m\le2$?
In particular, the case for $m=1$,
which corresponds to the original DQC1 model,
has been the long standing open problem.

In this paper, we show that if we can classically efficiently approximate
(exactly within a polynomial bit length 
or in the fully polynomial randomized approximation scheme (FPRAS)
with at most a constant error)
the output probability distribution of the DQC1 model,
then $\mbox{BQP}\subseteq\mbox{BPP}$.

Although the belief of
$\mbox{BPP}\subsetneq\mbox{BQP}$
is relatively less solid than the belief of P$\neq$NP
or that the polynomial hierarchy does
not collapse, researchers in quantum computing
believe $\mbox{BPP}\subsetneq\mbox{BQP}$
(for example, there is an oracle $A$ relative to which
$\mbox{BPP}^A\neq\mbox{BQP}^A$~\cite{BB}).
Therefore our results suggest that
DQC1 model can unlikely be classically efficiently simulated
in these senses.

If we replace the single pure qubit $|0\rangle$ of the DQC1 model with 
the completely-mixed qubit $I$,
the output probability distribution suddenly becomes easy to be classically
calculated.
Therefore, our results demonstrate 
``a power of a single qubit" in the DQC1 model.
Furthermore, it is known that some quantum circuits,
which look more complicated than the DQC1 model,
such as the stabilizer circuits~\cite{GK} and matchgates~\cite{match},
can be classically efficiently simulated.
These facts therefore suggest that the superficially naive DQC1 model
indeed hides some ability of doing complicated quantum computing.

Note that definitions of classical simulatability considered in this paper
are based on the calculation, and therefore stronger than the sampling,
which was used in Refs.~\cite{BJS,MFF}.
It is still an open problem whether DQC1$_m$ model for $m\le2$
is hard to be classically simulated in the sense of sampling.

\section{First Result}
Let 
$P(a):\{0,1\}\to[0,1]$
be the probability of obtaining the result $a\in\{0,1\}$ in the computational
basis measurement of the output qubit in the DQC1 model.
Let $P'(a)$ be an approximation of $P(a)$:
\begin{eqnarray*}
0\le P(a)-P'(a)\le\epsilon,
\end{eqnarray*}
where $0\le\epsilon\le\frac{1}{2^r}$
and $r$ is a polynomial of $n$ sufficiently larger than $n-1$.
If there exists a deterministic polynomial time
Turing machine that can calculate $P'(a)$,
then 
$\mbox{BQP}\subseteq\mbox{BPP}$.

\section{Proof}
Let $L$ be a language in the class 
BQP. This means that 
there exists a constant (error tolerance) $0<\delta<\frac{1}{2}$,
a uniform family of unitary operators $\{V_w\}$
acting on $|0\rangle^{\otimes n}$, where $n=poly(|w|)$, and
a specified single qubit output register $o\in\{0,1\}$ for the
$L$-membership decision problem such that
\begin{itemize}
\item[1.]
if $w\in L$ then $Prob(o=0)\ge 1-\delta$ and
\item[2.]
if $w\notin L$ then $Prob(o=0)\le \delta$.
\end{itemize}
By using the majority voting technique, $\delta$ can be
exponentially small~\cite{Barak}.

\begin{figure}[htbp]
\begin{center}
\includegraphics[width=0.25\textwidth]{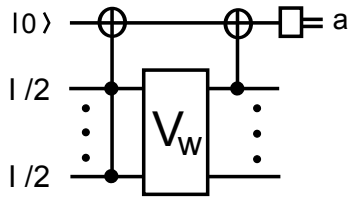}
\end{center}
\caption{
The DQC1 circuit created from the $n$-qubit Toffoli gate
and $V_w$.}
\label{DQC1circuit}
\end{figure}

For a unitary operator $V_w$,
we construct the DQC1 circuit of Fig.~\ref{DQC1circuit},
where the first $n$-qubit Toffoli gate is defined by
\begin{eqnarray*}
X\otimes |0\rangle\langle0|^{\otimes n}
+I\otimes (I^{\otimes n}-|0\rangle\langle0|^{\otimes n}),
\end{eqnarray*}
and $X\equiv|0\rangle\langle1|+|1\rangle\langle0|$
is the bit flip operator.

For this circuit, we obtain
\begin{eqnarray*}
P(a=0)=\frac{q}{2^{n-1}}
+\frac{1}{2}-\frac{1}{2^n},
\end{eqnarray*}
where
\begin{eqnarray*}
q\equiv\mbox{Tr}\Big[
\Big(|0\rangle\langle0|\otimes I^{\otimes (n-1)}\Big)
\Big(V_w|0\rangle\langle0|^{\otimes n}V_w^\dagger\Big)
\Big]
\end{eqnarray*}
is the probability of obtaining the positive result for
the BQP circuit.

By the assumption, there exists a deterministic polynomial time Turing machine
that can calculate $P'(a=0)$.
From the deterministic polynomial time Turing machine,
we can construct the probabilistic polynomial time Turing machine
that outputs $o\in\{0,1\}$
according to the probability 
\begin{eqnarray*}
Prob(o=0)&=&
2^{n-1}\Big[P'(a=0)-\frac{1}{2}+\frac{1}{2^n}\Big]\\
Prob(o=1)&=&
1-2^{n-1}\Big[P'(a=0)-\frac{1}{2}+\frac{1}{2^n}\Big].
\end{eqnarray*}

Then, the probabilistic polynomial time Turing machine outputs
$o=0$ in the following way:
\begin{itemize}
\item[1.]
if $w\in L$ then 
\begin{eqnarray*}
Prob(o=0)&=&
2^{n-1}\Big[P'(a=0)-\frac{1}{2}+\frac{1}{2^n}\Big]\\
&\ge&
2^{n-1}\Big[P(a=0)-\epsilon-\frac{1}{2}+\frac{1}{2^n}\Big]\\
&=&
2^{n-1}\Big[\frac{q}{2^{n-1}}-\epsilon\Big]\\
&=& q-\epsilon2^{n-1}\\
&\ge& (1-\delta)-\frac{1}{2^{r-(n-1)}},
\end{eqnarray*}
and
\item[2.]
if $w\notin L$ then 
\begin{eqnarray*}
Prob(o=0)&=&
2^{n-1}\Big[P'(a=0)-\frac{1}{2}+\frac{1}{2^n}\Big]\\
&\le&
2^{n-1}\Big[P(a=0)-\frac{1}{2}+\frac{1}{2^n}\Big]\\
&=&
q\\
&\le&\delta,
\end{eqnarray*}
\end{itemize}
and therefore $\mbox{BQP}\subseteq\mbox{BPP}$.

\section{Second Result}
Let us define
\begin{eqnarray*}
Q(a)\equiv P(a)-\frac{1}{2}.
\end{eqnarray*}
The fully polynomial randomized approximation
scheme (FPRAS) means that we can obtain an approximation
$Q'(a)$ of $Q(a)$, which satisfies
\begin{eqnarray*}
Prob\Big(\Big|Q(a)-Q'(a)\Big|\le\epsilon Q(a)\Big)\ge 1-\eta
\end{eqnarray*}
for given $\epsilon>0$ and $0<\eta<1$,
within $poly(\epsilon^{-1},\ln\eta^{-1},n)$ time.

Our second result is that if a classical computer can calculate $Q'(a)$
in the FPRAS with $\epsilon<\frac{1}{2}$ 
and $\eta<\frac{1}{2}$, and within $poly(n)$ time,
then $\mbox{BQP}\subseteq\mbox{BPP}$.
Note that 
$\eta=\frac{1}{4}$ is sufficient to reduce the failure probability
to an arbitrarily small value $\eta'$.

\section{Proof}
By the assumption, $Q'(a=0)$ can be calculated in the FPRAS.
Now we can construct the probabilistic polynomial time Turing machine
that outputs $o\in\{0,1\}$
according to the probability~\cite{largerthan1} 
\begin{eqnarray*}
Prob(o=0)&=&
2^{n-1}\Big[Q'(a=0)+\frac{1}{2^n}\Big]\\
Prob(o=1)&=&
1-2^{n-1}\Big[Q'(a=0)+\frac{1}{2^n}\Big].
\end{eqnarray*}

Then, the probabilistic polynomial time Turing machine outputs
$o=0$ in the following way: 
\begin{itemize}
\item[1.]
if $w\in L$ and $Q'(a=0)\ge Q(a=0)$ then 
\begin{eqnarray*}
Prob(o=0)&=&
2^{n-1}\Big[Q'(a=0)+\frac{1}{2^n}\Big]\\
&\ge& 2^{n-1}\Big[Q(a=0)+\frac{1}{2^n}\Big]\\
&=&q\\
&\ge&1-\delta.
\end{eqnarray*}
\item[2.]
if $w\in L$ and $Q'(a=0)\le Q(a=0)$ then~\cite{affect} 
\begin{eqnarray*}
Prob(o=0)&=&
2^{n-1}\Big[Q'(a=0)+\frac{1}{2^n}\Big]\\
&\ge&
2^{n-1}\Big[(1-\epsilon)Q(a=0)+\frac{1}{2^n}\Big]\\
&=&
(1-\epsilon)q+\frac{\epsilon}{2}\\
&\ge&
(1-\epsilon)(1-\delta)+\frac{\epsilon}{2}\\
&\ge&
(1-\epsilon)(1-\delta).
\end{eqnarray*}
\item[3.]
if $w\notin L$ and $Q'(a=0)\le Q(a=0)$ then 
\begin{eqnarray*}
Prob(o=0)&=&
2^{n-1}\Big[Q'(a=0)+\frac{1}{2^n}\Big]\\
&\le&
2^{n-1}\Big[Q(a=0)+\frac{1}{2^n}\Big]\\
&=&q\\
&\le&\delta.
\end{eqnarray*}
\item[4.]
if $w\notin L$ and $Q'(a=0)\ge Q(a=0)$ then~\cite{affect} 
\begin{eqnarray*}
Prob(o=0)&=&
2^{n-1}\Big[Q'(a=0)+\frac{1}{2^n}\Big]\\
&\le&
2^{n-1}\Big[(1+\epsilon)Q(a=0)+\frac{1}{2^n}\Big]\\
&=&(1+\epsilon)q-\frac{\epsilon}{2}\\
&\le&(1+\epsilon)\delta-\frac{\epsilon}{2}\\
&\le&(1+\epsilon)\delta.
\end{eqnarray*}
\end{itemize}
Therefore we conclude that $\mbox{BQP}\subseteq\mbox{BPP}$.

\section{Discussion}
In this paper, we have shown that the classical efficient approximation
(exactly within a polynomial bit length or in the FPRAS with at most
a constant error) of the output probability distribution
of the DQC1 model is impossible unless $\mbox{BQP}\subseteq\mbox{BPP}$.
Since it is believed that $\mbox{BPP}\subsetneq\mbox{BQP}$,
our results suggest that
it is unlikely that the DQC1 model can be simulated in these senses.

In Ref.~\cite{Datta3}, it was shown that the DQC1 model cannot be simulated
by using the tensor-network method, since the Schmidt rank increases exponentially.
Our first result can be considered as a generalization of their result:
whatever method is utilized, classical efficient simulation of the DQC1 model is impossible
(unless $\mbox{BQP}\subseteq\mbox{BPP}$).

As we have mentioned, our definitions of classical simulatability
are stronger than the sampling considered in Refs.~\cite{MFF,BJS}. 
It will be a future study to generalize our results to the
sampling.

\acknowledgements
TM is supported by the Tenure Track System by MEXT Japan
and KAKENHI 26730003 by JSPS.
TK is supported by KAKENHI 
26540002, 24106008, 24240001, 23246071
by JSPS.


\begin{thebibliography}{00}
\bibitem{largerthan1}
If $Q'(a=0)\le Q(a=0)$,
$2^{n-1}[Q'(a=0)+\frac{1}{2^n}]>1$
cannot be happen since
$2^{n-1}[Q'(a=0)+\frac{1}{2^n}]
\le 2^{n-1}[Q(a=0)+\frac{1}{2^n}]
=q$.
If $Q'(a=0)\ge Q(a=0)$,
and $2^{n-1}[Q'(a=0)+\frac{1}{2^n}]>1$,
we set $Prob(o=0)=1$,  
since in this case 
$2^{n-1}[Q'(a=0)+\frac{1}{2^n}]>1$
means
$q>\frac{2}{3}>\frac{1}{2}$.
\bibitem{affect}
Note that
this inequality is not satisfied with the small failure probability $\eta'$
due to the definition of FPRAS. However,
it does not affect the result.
\end{thebibliography}
\end{document}